\documentclass[11pt,twoside]{article}

\usepackage{asp2004}
\usepackage{epsf}
\usepackage{psfig}
\usepackage{lscape}

\newcommand{\nh}{n_{\rm H}}
\newcommand{\Nh}{N_{\rm H}}

\newcommand{\hm}{{\rm H}_2}

\newcommand{\LX}{L_{\rm X}}

\newcommand{\Msun}{M_{\odot}}
\newcommand{\Lsun}{L_{\odot}}
\newcommand{\pcc}{{\rm cm}^{-3}}

\newcommand{\psqcm}{{\rm cm}^{-2}}

\newcommand{\ps}{{\rm s}^{-1}}
\newcommand{\pyr}{{\rm yr}^{-1}}

\newcommand{\erg}{{\rm erg}}

\newcommand{\be}{\begin{equation}}
\newcommand{\ee}{\end{equation}}

\def\etal{\mbox{\it et al.\,}}

\begin{document}

\title{X-Ray Flares of Sun-Like Young Stellar Objects
and Their Effects on Protoplanetary Disks}

\author{A. E. Glassgold}
\affil{Astronomy Department, University of California, Berkeley, CA 94720}

\author{E. D. Feigelson}
\affil{Department of Astronomy and Astrophysics, Pennsylvania 
State University, University Park, PA 16802}

\author{T. Montmerle}
\affil{Laboratoire d'Astrophysique de Grenoble, 38041 Grenoble Cedex, France}

\author{S. Wolk}
\affil{Harvard-Smithsonian Center for Astrophysics, 60 Garden Street, 
Cambridge, MA 02138}

\begin{abstract} %%% Abstract to run on from here.

	Astronomical observations of flares from analogs of the early
	Sun have the potential to give critical insights into the high
	energy irradiation environment of protoplanetary disks.
	Solar-mass young stellar objects are significantly more X-ray
	luminous than the typical low-mass T Tauri star. They undergo
	frequent strong flaring on a several day time scale. Very
	powerful flares also occur, but on a longer time frame. The
	hard X-ray spectrum of these stars become even harder during
	flaring. The X-rays from these sun-like young stellar objects
	have the potential to ionize circumstellar material at a level
	greater than galactic cosmic rays out to distances $ \sim10^4
	$\,AU. Their characteristic hard
	spectra imply that, on encountering this material, they
	penetrate to fairly large surface densities of the order of 1
	g\,$\psqcm$ or more. Three specific illustrations are given
	of the effects of the X-rays: The physics and chemistry of the
	atmospheres of the inner accretion disks; the ionization level
	at the disk midplane, important for the viability of the
	magnetorotational instability; and the nuclear fluence in the
	irradiation zone just interior to the inner edge of the disk,
	important in local irradiation scenarios for producing the
	short-lived radionuclides found in meteorites.
 
\end{abstract}

%%% MAIN BODY OF TEXT GOES HERE. CONSULT "INSTRUCTIONS FOR AUTHORS USING
%%% LATEX2E MARKUP", SECTIONS 2.3-2.6 FOR HELP WITH EQUATIONS, FIGURES,
%%% AND TABLES.

\section{Introduction}

Ever since the EINSTEIN mission launched in 1978, it has been known
that young stellar objects (henceforth YSOs) are strong emitters of
X-rays. The X-ray emission exhibits large-amplitude flaring,
attributable to violent magnetic reconnection events similar to but
far more powerful than those seen on the contemporary Sun. During the
1990s, ROSAT and ASCA provided a wealth of new information on this
emission. With better angular resolution, ROSAT was able to measure
the soft X-rays from individual sources in nearby young clusters. ASCA
was able to detect hard X-rays, although its low angular resolution
led to some confusion as to their origin. Despite these limitations,
both observatories were extremely productive. In 1999, new-generation
X-ray observatories, CHANDRA and XMM/NEWTON, were launched. In
addition to the greater sensitivity associated with large mirrors and
detectors, these telescopes have the capability to measure X-rays over
a broad range of energies and with good spatial resolution. For
example, with its much improved spatial resolution, CHANDRA is ideally
suited to study the X-ray emission from the members of young stellar
clusters.

Many of the earlier results on the X-ray emission from YSOs were
summarized by some of the present authors in reviews written about the
time of the launch of CHANDRA and XMM/NEWTON. Feigelson \& Montmerle
(1999) gave a broad perspective on the observations and their
implications, and Glassgold, Feigelson, \& Montmerle (2000; henceforth
GFM00) focused on the effects of YSO X-rays on the circumstellar
environment. In this report, we update the observations with recent
results from CHANDRA, especially from a sample of solar-mass stars in
the Orion Nebula Cluster or ONC (Wolk et al.~2005) and address issues
of particular relevance to this workshop regarding the relation
between chondrites and the formation of the solar system. The ONC is
the nearest cluster (450\,pc distant) that has a full range of stellar
and sub-stellar masses, ranging from massive O stars in the Trapezium
to brown dwarfs. We also discuss the considerable progress that has
been made in understanding how X-rays affect the processes involved in
the formation of low-mass stars. We refer the reader to the two
above-mentioned reviews for additional details and earlier results.

\section{CHANDRA Observations of Solar-Mass YSOs}

Numerous papers have already been published from the five years of
observing YSOs with CHANDRA and XMM/NEWTON, usually based on exposures
lasting $\sim 50$\,ks. By combining two such observations for a total
of 85ks, CHANDRA was able to detect $\sim $1000 sources in the ONC
(Feigelson \etal 2002a), with $> 90 \%$ identified as YSOs. In an
unprecedented 800,000 s observation in early 2003 over 13 days, the
Chandra Orion Ultradeep Project (COUP) detected more than 1600 X-ray
sources in a $17\arcmin \times 17\arcmin$ image of the cluster
centered on the Trapezium stars in the Orion Nebula (Getman \etal
2005). A large fraction of these sources have been identified as pre
main-sequence objects. The long duration of the exposure offers a
unique opportunity to study the temporal behavior of YSO X-ray
emission. Many of the analyses of the COUP observations will be
published in an issue of the Astrophysical Journal Supplements in
2005. We restrict the present discussion to just two of the many
new observational findings: (1) a study of the flares of solar-mass
YSOs (Wolk \etal 2005), and (2) the detection of fluorescent
Fe line emission, presumably by circumstellar disk material (Tsujimoto
\etal 2005).

\subsection{X-ray Flares from Sun-Like YSOs}

Following the strategy of Feigelson et al.~(2002b), Wolk \etal (2005)
identified a small subset of the ONC X-ray sources as YSOs with masses
in the range $0.9-1.2\, \Msun$. The sample was defined with the help
of the stellar evolutionary tracks calculated by Siess \etal (2000)
that give approximate ages for the sample members. Most are in the
range from $1-5$\,Ma, with a median age of $\sim 2$\,Ma. Figure 1
shows a 2MASS (Two Micron All Sky Survey) infrared photo of the
central $12\arcmin$ of the ONC, corresponding to a scale 1.57\,pc at
Orion's distance of 450\,pc. The COUP solar-mass stars, observed by
CHANDRA with a spatial resolution of $\sim 1 \arcsec$, are shown as
green circles.

\begin{figure}[h]
\plotone{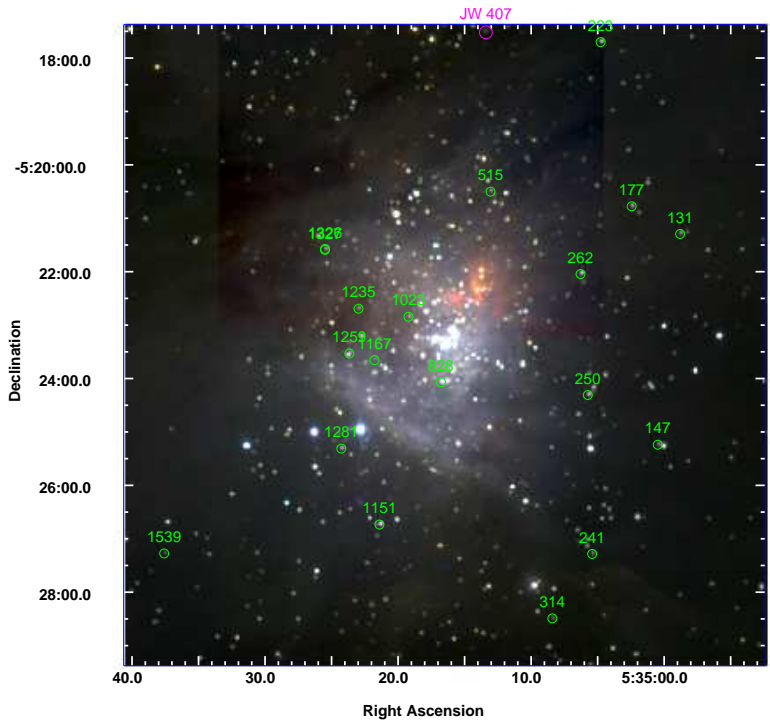}
\caption{The central part of the Orion Nebula cluster with the 
COUP solar-mass stars labeled by (green) circles. \label{coup}}
\end{figure}

The COUP observations confirm previous CHANDRA results (anticipated by
the observations of earlier X-ray satellites) that solar-mass YSOs are
strong emitters of hard as well as soft X-rays. The ``characteristic''
X-ray luminosity for the sample has a median value $\LX \simeq
10^{30}\, \erg \,\ps$and a distribution that has a spread $ \pm 10$
about the median~\footnote {Instead of quiescent luminosity, Wolk
\etal (2005) define a characteristic luminosity in terms of Bayesian
blocks of the X-ray light curve when the YSO emission is essentially
constant, which is about 75\% of the time.}. This absolute level is
larger than for most T Tauri stars, which have lower masses. Preibisch
\etal (2005) show that, for the COUP data on the ONC cluster, the
X-ray luminosity increases significantly with stellar mass, roughly as
$\LX \propto M^{1.5}$ for $0.1 < M/\Msun < 2$, where $\LX$ is the
present luminosity of the Sun. The exact dependence is a function of
the pre main-sequence evolutionary tracks used to determine age and
mass.

Figure 2 shows an example of the spectrum of a solar-mass YSO. This
particular source (\#567) is a $1.2\,\Msun$, K-type, moderately
extincted YSO that underwent three strong flares during the COUP
observation, as discussed below. It illustrates well the generic
features of solar-mass YSO X-ray spectra. First, the soft X-rays
(roughly those below 1 keV) are strongly absorbed by interstellar
matter along the line-of-sight, an effect due to the strong decrease
in the X-ray absorption cross section with X-ray energy, $\sim
E^{-2.65}$. Second, the flux level of the hard X-rays (above 1 keV) is
high. Because of the energy dependence of the absorption cross
section, a 1-keV X-ray has an absorption length $\sim 10^{22}\psqcm$,
whereas a 5-keV X-ray has an absorption length $>
10^{24}\psqcm$. These facts are relevant for the discussion to follow
of the environmental effects of the X-rays: solar-mass YSOs produce
X-rays that can penetrate through relatively large surface densities
of matter. This capability is enhanced during flares when the spectrum
hardens.

\begin{figure}[h]
\plotfiddle{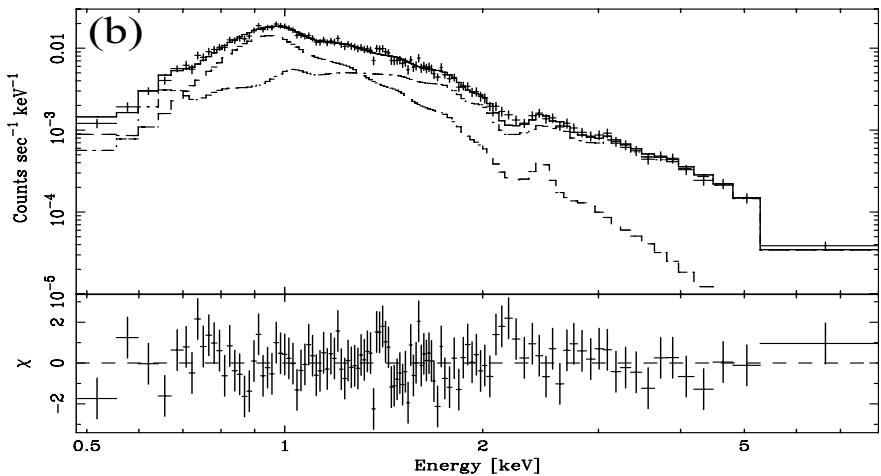}{2.5in}{0.}{50.}{40.}{-170.}{0}
\caption{X-ray energy spectrum of COUP source \# 567 from Fig.~6b 
of Getman \etal (2005). The spectrum is fit (heavy solid line) by
a conventional two-temperature model with $kT_1 =0.8$\,keV and 
$kT_2 = 3.1$\,keV (lighter lines). The residuals present evidence
of Ne-Fe anomalies often seen in strong flares. The sub-keV emission
is absorbed out by intervening material with a column density, 
$\Nh = 2 \times 10^{21} \psqcm $.\label{spectrum}}
\end{figure}

Figure 3 shows the light curves of two of the COUP solar-mass stars in
the the Wolk \etal (2005) sample. They are reasonably typical examples
in that they show several flares during the 10-day observing period,
where a flare is defined as a statistical significant large and rapid
rise above the characteristic level. A noteworthy aspect of Figure 3
is that there is evidence here for both short-duration ($\la
1$\,hour) and long-duration flares ($\ga 1 $ \,hour). It should
also be mentioned that the observations are limited in their
capability to detect all but the strongest short-duration flares.

\begin{figure}[h]
\plotfiddle{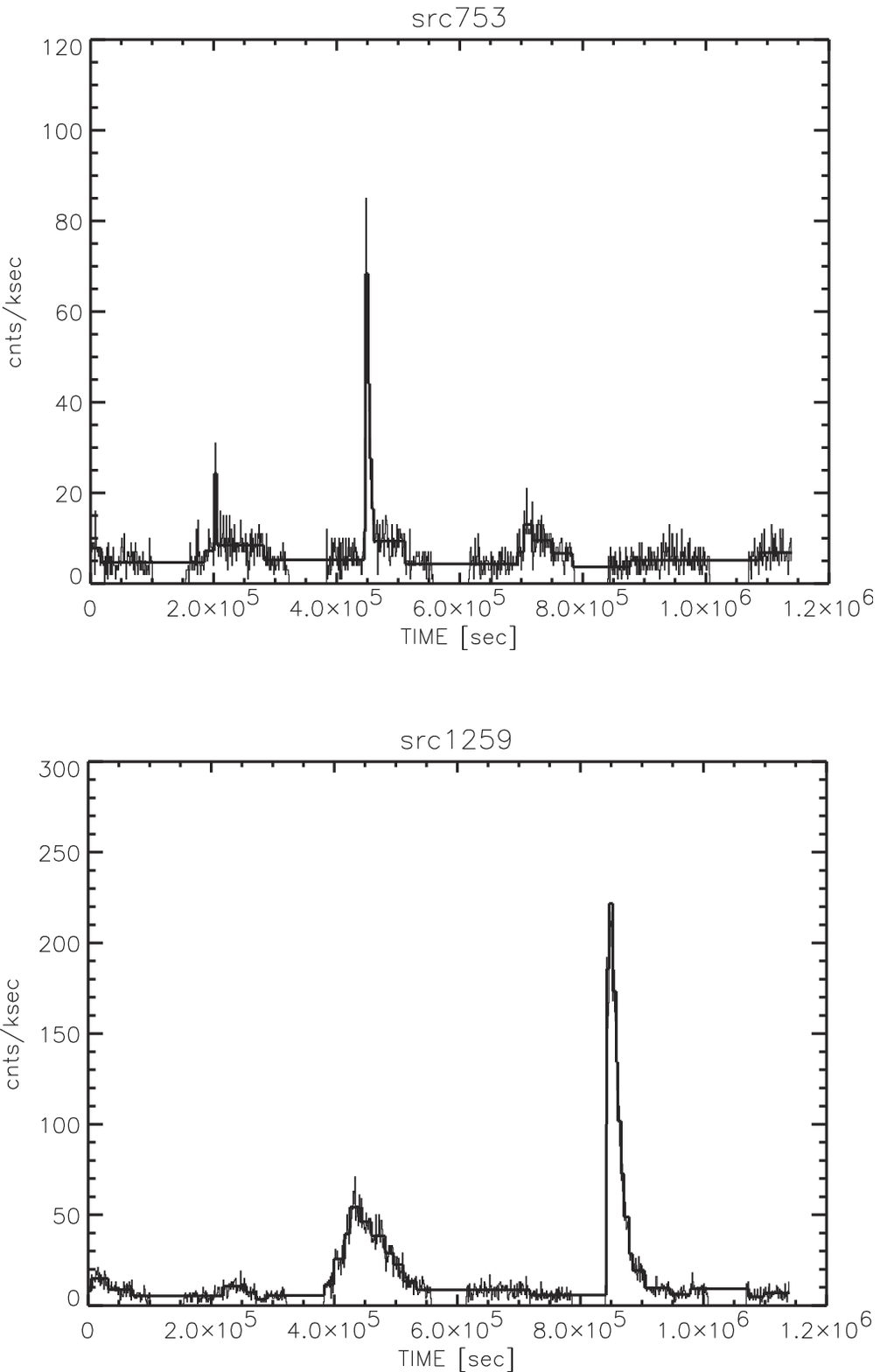}{5.5in}{0.}{100.}{90.}{-150.}{0}
\caption{X-ray light curves for two COUP sources showing several
types of flares. The second flare for source \#1259 (bottom) has a peak
luminosity of $0.1 \Lsun$, one of the most powerful ever
observed from a YSO. \label{flares}}
\end{figure}

Another consequence of the flare analysis of Wolk \etal is that the
flare frequency, one every several days, is significantly smaller than
the flare frequency obtained by Feigelson \etal (2002b) for a sample
of solar-mass stars selected in a different way. The YSOs in the COUP
sample of Wolk \etal have a median peak luminosity of $\LX = 10^{31}
\erg \ps$. The peak of the second flare in source \# 1259 at the
bottom of Figure 3 is $\LX = 0.1 \Lsun$. This is one of the most
luminous YSO X-ray flares observed so far. Many of the flares seen by
COUP do not have the classical form consisting of a rapid rise and a
simple exponential-like decay.  The first flare in source \# 1259 in
Figure 3 is a good example. Perhaps it involved a sequence of events
initiated by the first flare.

To summarize, strong flares are common in the COUP YSOs with solar
masses. On average, they occur on a time scale of several
days. Gigantic flares, such as the event in source \# 1259 that
reached $\LX = 0.1 \Lsun$, are rarer and occur on time scales of the
order of one year, or perhaps are restricted to a subset of solar-mass
YSOs. A last important fact about this flare sample is the strong
evidence that the spectra become significantly harder during a flare
(Wolk \etal 2005).

\subsection{Fluorescent Fe Line Emission}

Another relevant result from COUP is the detection by CHANDRA of
fluorescent Fe-line emission near 6.4\,keV from a number of YSOs
(Tsujimoto \etal 2005). These authors interpret their observations as
indicative of the absorption of hard photons by circumstellar
material, probably the disk. The K-shell threshold for Fe ions in low
ionization states occurs near 6.4\,keV and increases to 6.9 \,keV for
almost fully ionized ions. Not only do CHANDRA spectra of YSOs contain
photons ranging up to and beyond this energy range, they often show an
unresolved emission feature near 6.9\,keV appropriate to the coronal
temperatures of the X-ray source. These hard photons are absorbed by
circumstellar Fe in low-ionization states which then decay by the
Auger process and by fluorescent X-ray emission. Although the
probability for fluorescent decay is small for the light elements, it
is quite significant for Fe, $\sim 25\%$.

Figure 4 shows the spectrum of one of the seven COUP sources for which
the 6.4\,keV fluorescent line was detected by Tsujimoto \etal
(2005). These sources were found in a control sample of 123 candidates
with large hard X-ray emissivities. They all have near infrared colors
indicative of disks, flares with amplitudes larger and spectra harder
than average for the control sample, and more absorption than average.
Tsujimoto \etal argue as follows that the fluorescence originates in a
flattened disk. First, it cannot originate in the stellar photosphere,
since otherwise it would have been seen in all sufficiently bright
COUP sources, rather than only in a few embedded YSOs with strong
infrared excesses. Second, it cannot be fluorescence from
line-of-sight material, since the column densities (known for each
source from the soft X-ray absorption) are too small by a factor of
100. Fluorescence from a spherical envelope can be similarly
rejected. The remaining alternative is fluorescence from circumstellar
material, i.e., a flattened concentration of material that is
efficiently illuminated by the X-ray source but does not obscure the
line-of-sight, e.g., a disk~\footnote{Of course other circumstellar
material can scatter hard X-ray photons, e.g., the wind and funnel
accretion column.}.  Other studies support this conclusion.  The
6.4\,keV fluorescent line had been previously seen with CHANDRA in a
few protostars (e.g. YLW 16A in the Ophiuchi cloud; Imanishi et
al.~2001).  Another COUP Study by Kastner \etal (2005) demonstrates
the absorption of X-rays in disks by a correlation between soft X-ray
absorption and the inclination of proplyds imaged with the Hubble
Space Telescope.  Altogether, these observations provide important
evidence that YSO X-rays do, at least in some cases, directly
irradiate protoplanetary disks.

\begin{figure}[h]
\plotone{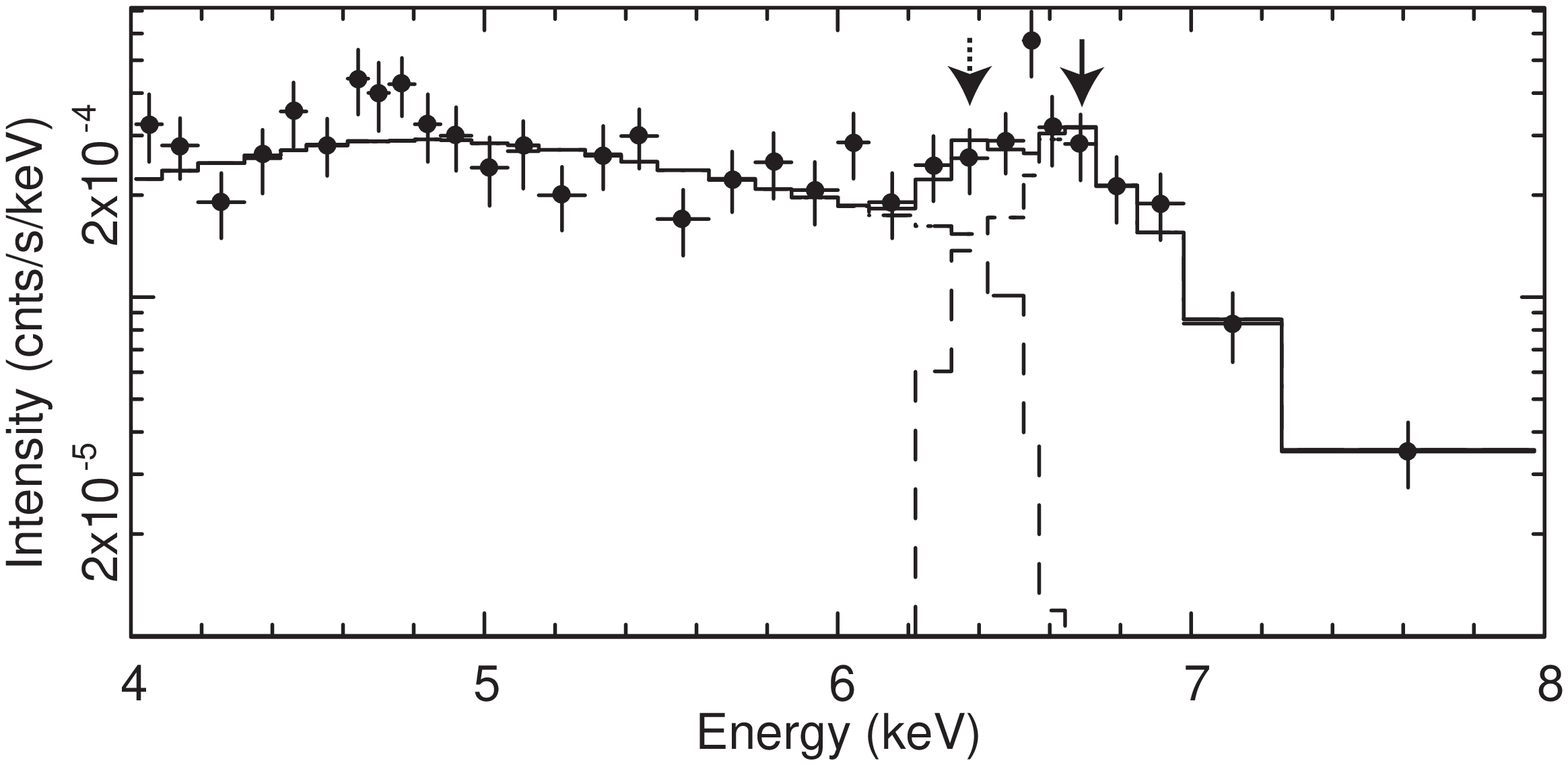}
\caption{Hard X-ray spectrum of a COUP source with a 6.4\,keV
feature indicative of scattering by a circumstellar disk (Tsujimoto
\etal 2005, as discussed in \S 2.2). The crosses are the data. A
best-fit model is represented by solid steps and Gaussian line
components by dashed steps. The 6.4 keV (low-ionization Fe) and 6.7
keV (high ionization Fe) features are indicated by solid and broken
arrows, respectively. \label{tsuji}}
\end{figure}

\section{X-ray Interactions} 

X-rays from a sun-like star in the ONC can affect the physical
conditions in its immediate environment, as discussed in the earlier
review by GFM00. In gauging these effects, one has to keep in mind the
possible competition from other ionizing sources such as UV photons
and cosmic rays. In a large moderately dense young cluster like the
ONC, with a more or less complete range of stars from the brown dwarf
limit up to O stars, the intra-cluster radiation field can also play
an important role, especially in the neighborhood of the Trapezium
stars. Sufficiently close to any given star, however, the X-ray
emission from that star (and its companion if any) will dominate. This
is probably true as a first approximation, even in the presence of the
strong UV radiation produced by the accretion shock on the stellar
surface, partly because of the greater absorption of the far UV
(912-1100\,\AA) by small dust particles, atomic C, CO and $\hm$. In
addition, satellite observations show that this part of the UV
spectrum of T Tauri stars is relatively weak (e.g., Valenti \etal
2000; Johns-Krull 2000).

Focusing on a parcel of circumstellar matter, the effects of the YSO
X-rays arise from the electron cascade initiated by the absorption of
a moderately hard (keV) YSO photon by a K or L shell electron of a
low-abundance heavy atom or ion. The absorption process itself leaves
an excited ion, whose main de-excitation mode is the ejection of more
(Auger) electrons with energies of the same order of magnitude as the
initial threshold. The energy of these first electrons, photo and
Auger, is then degraded in a series of collisions that excite and
ionize the main constituents of the cosmic gas, H and He, and also by
Coulomb collisions with the ambient electrons. The number of
collisions is large, since only 35-40\,eV is needed to make an ion
pair in a low-ionization gas with cosmic-like abundances. From the
point of view of the macroscopic physical properties of the
circumstellar matter, the main role of the X-rays is to affect its
ionization and chemical properties and its thermal state via
electronic collisions. The physical and chemical effects generated by
X-ray irradiation are discussed by Maloney, Hollenbach, and Tielens
(1996) with extensive references to earlier work.

In order to gain an appreciation of the magnitude of these effects, we
calculate the ionization rate at a radial distance of 1\,AU from a
YSO, ignoring attenuation and scattering to be discussed below. We
update the previous discussion of GFM00 by following Glassgold,
Najita, \& Igea 2004 (henceforth GNI04). Using the median
characteristic X-ray
luminosity from Wolk \etal (2005) as a scale factor, we obtain,
\begin{equation}
\label{ionizationrate}
\zeta = {6\times 10^{-9}} \ps \, 
(\frac{L_{char}}{2 \times 10^{30} \erg \,\ps})\, 
(\frac{\rm AU}{r})^2.
\end{equation}
This number is larger than the formula given by GFM00 because we are
now using (1) a nominal luminosity appropriate to a sun-like YSO,
rather than $\LX = 10^{29}\,\erg \, \ps$ and (2) a smaller low-energy
cut-off to the spectrum, 0.1\,keV, instead of 1\,keV. It is worth
repeating that the above X-ray ray ionization rate at 1\,AU is eight
orders of magnitude larger than the ionization rate due to galactic
cosmic rays in the solar neighborhood, $\zeta_{\rm CR} = 5 \times
10^{-17} \, \ps$. Thus, {\it if} stellar X-rays and galactic cosmic
rays are the only external ionization source, X-rays can in principle
dominate out to $10^4\,{\rm AU} = 0.05$\,pc, taking inverse-square
dilution into account but ignoring attenuation and scattering of both
radiations, which in any real application need to be included.

Equation~\ref{ionizationrate} is essentially the same as Eq.~(4) of
GFM00, except that the attenuation factor has been dropped and new
choices made for the parameters. Since the X-ray luminosity of YSOs
can change greatly from one YSO to another, as do their other
properties, the value of $\LX$ in Eq.~\ref{ionizationrate} needs to be
chosen to fit the particular application under consideration.
Furthermore, the effect of the large X-ray ionization rate for
solar-mass stars is mediated in applications to thermal and chemical
phenomena. For example, $\zeta$ enters into the ionization equation
through the combination $\zeta / \nh$, so that the disk density
distribution is an important factor. Furthermore, the actual
ionization fraction usually depends on a fractional root of $\zeta /
\nh$, typically between 1/2 and 1/3 at intermediate densities and
probably even smaller at higher densities. Thus variations in $\LX$ of
the order of 10 may change the ionization by only a factor of a few.

The above comparison of the X-ray and cosmic ray ionization rates is
relevant for thick dense regions; otherwise far UV radiation (FUV)
becomes important. For example, let us compare the above X-ray
ionization rate at 1\,AU for a solar-mass YSO with the rates at which
the interstellar UV radiation field ionizes neutral carbon ($2 \times
10^{-10}\, \ps$) or dissociates CO and $\hm$ ($\sim 5 \times
10^{-11}\, \ps$). FUV can dominate X-ray destruction under some
circumstances, e.g., at larger distances or for smaller X-ray
luminosities. On the other hand, FUV radiation is, almost by
definition, unable to ionize hydrogen and helium. Close enough to a
YSO, the X-rays tend to dominate.
  
Because the X-ray spectrum includes a wide range of photon energies,
the flux decreases less rapidly with column density than exponentially
(Krolik and Kallman 1983). Glassgold, Najita and Igea (1997,
henceforth GNI97) gave an improved theory, still in closed form, and
Igea and Glassgold (2001) included the effects of scattering in a
Monte Carlo calculation of disk ionization, to be discussed below in
more detail. Although the formulae in GNI97 remain valid, it is not
much more difficult in applications to numerically calculate the total
attenuation along the line of sight from look-up formulae or tables of
the absorption cross section, as done by GNI04. The most up to date
account of the X-ray absorption cross sections is Wilms, Allen and
McCray (2000).
 
In the rest of this section, we consider regions close to the YSO and
assume that the most important ionizing radiation comes from the
direct X-ray emission of the star, assumed for simplicity to be
single. Star formation is a complex time-dependent process involving
several dynamical components, and our understanding of the effects of
X-rays is far from complete. We discuss three specific topics of
current interest for YSOs of roughly solar mass: the atmospheres of
the inner regions of YSO accretion disks, the ionization level near
the mid-plane of the disk, and the irradiation zone interior to the
disk proper. Extensive studies have also been carried out on the jets
of active but revealed YSOs (Shang \etal 2002, henceforth
SGSL02). This theory has recently been extended to the radio continuum
emission from younger and more embedded sources (Shang \etal
2004). The effects of X-rays on the chemical abundances of disks at
large radii have been studied by many authors, e.g., Aikawa \& Herbst
(1999, 2001), Markwick \etal (2002), and Gorti \& Hollenbach (2004),
the last with specific emphasis on mid-infrared diagnostics relevant
for the Spitzer Space Telescope.

Here we focus on: (1) a novel application to the atmospheres of disks
at small radii, (2) the ionization of the disk mid-plane, fundamental
for understanding disk viscosity, and (3) the connection between X-ray
and nuclear irradiation of material just inside the inner radius of
the disk, a situation directly relevant for meteoritics. The effects
of X-rays on the physical properties of the funnel accretion flow have
not yet been investigated.

\subsection{Disk Atmospheres}

Although there have been numerous theoretical and observational
studies of {\it dust} in disks, much less attention has been paid to
the {\it gas}. Most theoretical work on the thermal properties of
disks has assumed that the gas and the dust are at the same
temperature. However, in the study of how X-rays affect the upper
atmospheres of protoplanetary disks, GNI04 found that this is
definitely not the case away from the mid-plane (above a few scale
heights). Near the midplane, the densities are high enough for the
dust and gas to interact strongly thermally.  Going away from the
mid-plane, the disk becomes optically thin to the stellar optical and
infrared radiation, and the temperature of the (small) grains will
rise, as does the still closely-coupled gas temperature. However, at
still higher altitudes, the gas responds to the strong X-ray flux as
the attenuation of the X-ray decreases. Its temperature rises to very
large values, much higher than that of the dust. To reach this
conclusion, GNI04 performed a self-consistent thermal-chemical
calculation to determine the abundances of the gaseous ions, atoms,
and molecules that are important for cooling the gas through line
radiation. One simplification is the use of the gas density and dust
temperature obtained by D'Alessio \etal (1999) in a thermal
calculation for a generic T Tauri disk. In principle, the fact that
the dust and gas are not perfectly coupled implies that a more general
two-fluid model is needed. This is important for the observed
properties of the dust, such as the spectral energy distribution,
since the vertical variation of the atmosphere, for example the degree
of flaring, is strongly affected by the gas temperature and pressure.

Figure 5 shows some of these results for the atmosphere of a T Tauri
disk at a radial distance of 1\,AU starting from the D'Alessio \etal
(1999) model with an accretion rate of $10^{-8} \, \Msun \, \pyr$.
Temperature is plotted against perpendicular column density measured
from the top of the atmosphere~\footnote{The calculations start at a
height where the hydrogen gas density is $10^{7}\, \pcc$.}. The dashed
line shows the dust temperature rising above its midplane value due to
the increase in illumination with height of small dust grains by the
stellar radiation. The dust temperature undergoes a modest inversion,
going from about 140\,K at midplane to 450\,K at the top of the
disk. At a vertical column $\sim 10^{22} \psqcm$, the gas temperature
begins to depart from that of the dust and, due to surface heating,
eventually manifests a very large temperature inversion. In the top
layers fully exposed to YSO X-rays, the gas temperature is of order
5000\,K. In between the top of the atmosphere and the mid-plane, there
is a warm transition region (500-2000\,K) with the interesting
chemical property that is it is mainly atomic hydrogen but has carbon
fully associated into molecules. In such a mixture, the CO
rotation-vibration transitions are easily excited because of the large
collisional excitation cross sections for H + CO inelastic scattering
(much larger than for molecular hydrogen). The occurrence of an X-ray
heated hot layer on top of a cold mid-plane layer has also been
obtained by Alexander, Clarke and Pringle (2004).

\begin{figure}
\plotfiddle{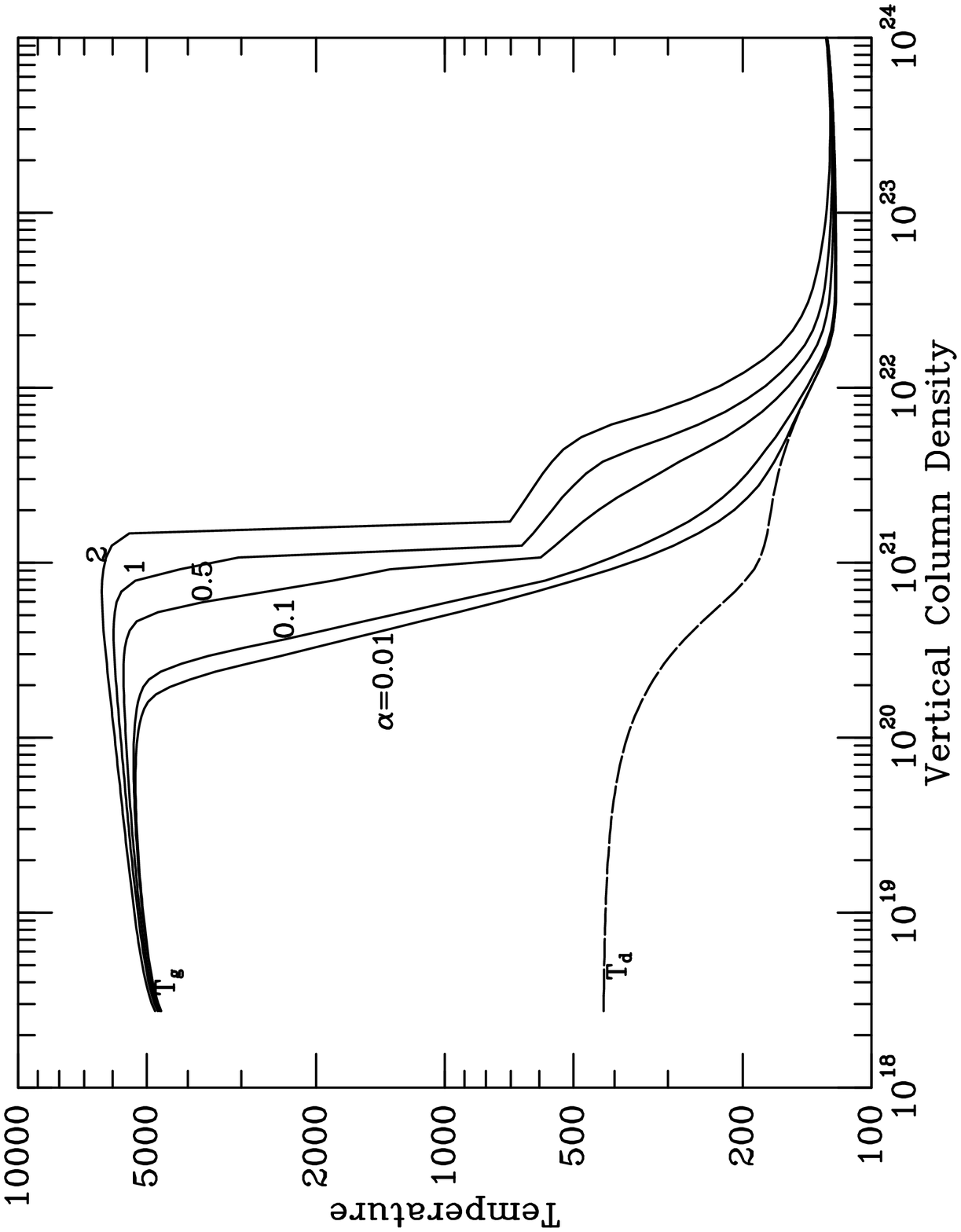}{3.0in}{270.0}{50.}{40.}{-185.}{230}
\caption{Temperature profiles from GNI04 for the surface of the
protoplanetary disk atmosphere discussed in \S 3.1 of the text. The
radial distance is 1\,AU, and the mass-loss rate is $10^{-8} \Msun
\pyr $. The dashed line is the dust temperature of D'Alessio \etal
(1999). The solid curves are gas temperature profiles labeled by the
coefficient $\alpha$ in Eq.~\ref{accheat}. The choice $\alpha=0.01$
closely follows the limiting case of pure X-ray heating. The abscissa
is vertical column density in cm$^{-2}$, and the ordinate is temperature 
in degrees K. \label{flows}}
\end{figure}

The appearance of several temperature curves in Figure 5 reflects the
fact that GNI04 considered other surface heating mechanisms in
addition to X-ray heating. In particular, they investigated the
potential role of heating by the YSO wind interacting with the upper
layers of the disk and heating by mechanical processes associated with
the outward transfer of angular momentum. Although heating from these
processes surely occurs, the theoretical understanding of both is
incomplete, and GNI04 treated them phenomenologically. For example,
the total wind energy incident on the disk can be calculated for any
wind model, but the depth and other properties of the resultant
turbulent mixing layer are poorly constrained. In the case of viscous
heating associated with angular momentum transport, the most popular
mechanism is the magnetic rotational instability (MRI) introduced to
disk physics by Balbus \& Hawley (1991; see also the review by Stone
\etal 2000). It leads to local viscous heating of the usual form,
\be
\label{accheat}
\Gamma_{\rm acc} = \frac{9}{4} \alpha \rho c^2 \Omega, 
\ee 
where $\rho$ is the mass density, $c$ is the isothermal sound speed,
$\Omega$ is the angular rotation speed, and $\alpha$ is a
characteristic parameter measuring the degree of magnetic turbulence.
GNI04 argued on the basis of simulations by Miller and Stone (2000)
that midplane turbulence generated Alfv\'en waves that, on reaching
the diffuse surface regions, produce shocks and heating. GNI04 regard
$\alpha$ as a phenomenological parameter to be determined by
appropriate observations of disk atmospheres. In practice, they lump
both surface heating mechanisms together since, by dimensional
arguments, they may have similar forms.  This parameter $\alpha$
labels the gas temperature in Figure 5.

	According to Figure 5, there is a narrow transition region
	with gas temperatures in the range 500-2000\,K, that lies
	between the chromospheric-like temperatures at high altitudes
	and the relatively cool temperature at mid-plane. The location
	and thickness of this layer depends on the strength of the
	surface heating. The curves illustrate this dependence in the
	case of a T Tauri disk at $r=1$\,AU. For $\alpha = 0.01$,
	X-ray heating dominates this region, whereas for $\alpha >
	0.1$, mechanical heating is paramount. When these ideas are
	applied to the more actively-accreting T Tauri stars,
	mechanical heating is required to explain the strength of the
	CO fundamental emission observed by Najita, Carr \& Mathieu
	(2003).  GNI04 concluded that some form of surface heating,
	such as wind-disk interactions or accretion, is required to
	explain the observations. This is rather similar to the
	conclusion reached by SGSL02 and SLSG04 in their study of the
	jets of large accretion-rate YSOs, where X-rays were found to
	be more important for ionization than for heating. The hot on
	cold layer model with a warm interface offers new
	opportunities for diagnostic probes of the gas in
	protoplanetary disk atmospheres and for verifying that YSO
	X-rays are effective in determining their physical and
	chemical properties.

\subsection{Near Mid-Plane Ionization and the MRI}

	The crucial role of the ionization level in disk accretion via
	the MRI was pointed out by Gammie (1996). The physical reason
	is that collisional coupling between electrons and neutrals is
	required to transfer the turbulence in the magnetic field to
	the overwhelmingly neutral material of the disk. Gammie found
	that galactic cosmic rays could not penetrate beyond a thin
	surface layer of the disk at several AU. He suggested that
	accretion only occurs in the surface of the inner disk (the
	"active region") and not in the much thicker mid-plane region
	(the ``dead zone'') where the ionization level is too small to
	mediate the MRI.  Glassgold, Najita \& Igea (1997) argued that
	the galactic cosmic rays could not penetrate the surface
	region of the inner disk because they are blown away by the
	stellar wind, much as the solar wind does with solar energetic
	particles.  They showed that YSO X-rays could do almost as
	good a job as cosmic rays in ionizing surface regions, thus
	preserving the layered accretion model of the MRI for
	YSOs~\footnote{GNI04 also suggested that Gammie's dead zone
	might provide a good environment for the formation of
	planets.}. Igea \& Glassgold (1999) supported this conclusion
	with a Monte Carlo calculation of X-ray transport through
	disks, demonstrating that scattering plays an important role
	in the MRI by extending the active surface layer to column
	densities greater than $10^{25} \, \psqcm$.

	The result of this first work is that the theory of disk
	ionization and chemistry is crucial for understanding the role
	of the MRI for YSO disk accretion and possibly for planet
	formation. These challenges have been taken up by several
	groups, largely in the context of layered accretion e.g., Sano
	\etal (2000), Fromang \etal (2002), Semenov \etal (2004), Kunz
	\& Balbus (2004), Desch (2004) and Matsumura \& Pudritz (2003,
	2005). Fromang \etal discussed some of the issues that can
	have a significant affect on the size of the dead zone:
	differences in the disk model, such as a Hayashi disk or a
	standard $\alpha$-disk; temporal evolution of the disk; the
	role of a small abundance of heavy atoms that recombine mainly
	radiatively; and the value of the magnetic Reynolds
	number. Sano \etal (2000) explored the part played by small
	dust grains in reducing the electron fraction when it becomes
	as small as the very small abundance of dust grains. They
	showed that the dead zone decreases and eventually vanishes as
	the grain size increases or as sedimentation towards the
	mid-plane proceeds. Kunz \& Balbus (2004) and Desch (2004)
	have addressed the growth and saturation of the instability
	itself, in particular the effects of initial assumptions about
	the magnetic field and the role of various diffusive
	processes. Matsumura \& Pudritz (2003, 2005) have investigated
	some of the implications of layered accretion for planet
	formation. For further discussion of the role of the MRI in
	the formation of the Sun, see the article by Gammie and
	Johnson, this volume.

	Semenov \etal (2004) have employed the most complete chemical
model to calculate the mid-plane ionization. Like Fromang \etal, they
include X-ray ionization following GNI97 and ignore scattering; they
also employ too small an X-ray ionization rate for Sun-like stars.
They include the effects of galactic cosmic rays at something close to
a standard rate ($\zeta_{\rm CR} \sim 10^{-17} \ps$) and UV radiation
from both the mean galactic field (following Draine 1978) and from the
star itself. Following earlier chemical models oriented towards the
outer regions of disks (e.g., Willacy \& Langer 2000; Aikawa \& Herbst
1999, 2001; Markwick \etal 2002), Semenov \etal adopt a value for the
{\it stellar} UV radiation field that is $10^4$ times larger than
galactic at a distance of 100\,AU. This choice can be traced back to
early IUE measurements of the stellar UV beyond 1400~\AA \, for
several T Tauri stars (Herbig and Goodrich 1986).  Although the UV
flux from T Tauri stars covers a range of values and is undoubtedly
time-variable, more detailed studies using both IUE (e.g., Valenti
\etal 2000; Johns-Krull \etal 2000) and FUSE (e.g., Wilkinson \etal
2002; Bergin \etal 2003) indicate that it decreases going into the FUV
domain and has a typical value, $\sim 10^{-15} \erg \, \psqcm \ps$ \AA
$^{-1}$, that is significantly smaller than earlier estimates. The FUV
band is crucial because this is where atomic C can be photoionized and
$\hm$ and CO photodissociated.
  
	Confining ourselves to considerations of the inner disk, the
issue of whether stellar FUV or X-ray radiation dominates is important
for the ionization and chemistry of protoplanetary disks because of
the vast difference in the energy of the photons. The most direct
consequence is that FUV photons cannot ionize H, and thus the
abundance of carbon provides an upper limit to the ionization level
produced by the photoionization of heavy atoms, $\sim 10^{-4}$ -
$10^{-3}$. Next, FUV photons are absorbed much more readily than
X-rays, although this depends on the size distribution of the dust
grains, i.e, on grain growth and sedimentation. If we use realistic
numbers for the FUV and X-ray luminosities of T Tauri stars, we find
that $L_{\rm FUV} \sim \LX$. The rates used by Semenov \etal (2004)
and many other modelers of disk chemistry correspond to $\LX \ll
L_{\rm FUV} $. Therefore, the important study of Semenov \etal needs
to be extended to the relevant range in the $L_{\rm FUV} - \LX$
parameter space. More generally, the disk parameters used to explore
disk chemistry, i.e., accretion rate, FUV and X-ray luminosity, etc.,
should correspond as much as possible to the actual T Tauri stars for
which theory and observations are to be compared. Likewise the
complete neglect of the FUV in GNI04, initially justified by the large
difference between the X-ray and FUV attenuation factors, should also
be remedied. The simultaneous treatment of the transfer of the FUV and
X-ray radiation is complicated by the fact that the FUV
photo-destruction of both $\hm$ and CO proceed by line absorption and
thus involves highly nonlinear line self-shielding. The importance of
solving this problem is highlighted by the suggestion that oxygen
isotope anomalies in meteorites are determined by selective
photodissociation of CO determined by line self shielding (Clayton
2002; Lyons \& Young 2004; Yurimoto \& Kuramoto 2004; Yin 2004; and Krot
2005, these proceedings).

The level of ionization in the mid-plane of YSO accretion disks is
affected by many dynamical and microphysical processes. At present,
uncertainties in these processes preclude a definitive quantitative
evaluation of the ionized regions of protoplanetary disks.  But some
overall characteristics are clear.  Stellar X-rays and FUV emission
are significant sources of heating, ionization, and chemical activity
in the inner regions of protoplanetary disks.  Although the X-rays
penetrate deeply towards the midplane at moderately large radii, they
ionize only a shallow region above a neutral ``dead'' zone at small
radii.

\subsection{Stellar Energetic Particles and Short-Lived Radionuclides}

	In calculating the production of short-lived radionuclides in
the region near and inside the inner disk or co-rotation radius, Lee
et al.~(1998) invoked ROSAT and ASCA observations of soft and hard YSO
X-rays to estimate the fluence of nuclear particles. They converted
X-ray to stellar energetic-particle fluxes using observations of the
contemporary active Sun. A related calculation was done by Feigelson,
Garmire, \& Pravdo (2002b) using CHANDRA observations. They estimated
that the particle fluxes from active YSOs were $\sim 10^5$ more
powerful than in the active Sun. This number reflects the fact that
YSO flares are more powerful and more frequent than in the
contemporary Sun. They also took into account that the distribution in
energy of solar energetic particles is shallower than that of the
X-rays. This factor of $10^5$ enhancement in flare particle fluences
is sufficient to produce by spallation reactions several important
anomalies in the abundances of short-lived radionuclides found in
meteoritic CAIs (Lee \etal 1998, Goswami \etal 2001, Gounelle \etal
2001, Marhas \etal 2003, Leya \etal 2003).

	The COUP observations of solar-mass YSOs can be used to make
similar estimates of the particle fluence in the reconnection ring.
>From the recent COUP study of solar-mass Orion stars (Wolk et al.~
2005) discussed in \S 2.1, we take the mean characteristic luminosity
as $10^{30}$\,erg\,s$^{-1}$, and the mean flare luminosity, duration,
and repetition times as $5 \times 10^{30}$\,erg\,s$^{-1}$, $10^5$\,s,
and 4-8\,d, respectively, and estimate the fluence at a distance $0.75
R_x$ (with $R_x = 0.05$\,AU, the x-point or co-rotation radius) over
10 years to be,
\begin{equation}
{\cal F}_{\rm X}(10\,{\rm yr}) = 2 \times 10^{15} {\rm erg} \, \psqcm.
\end{equation}
If we convert from X-ray to proton fluence (for energies greater than
10 MeV) using 0.1 as the conversion factor (e.g., Lee et al.~1998), we get essentially the same result as these authors,
\begin{equation}
{\cal F}_p(10\,{\rm yr}) = 2 \times 10^{14} {\rm erg}\, \psqcm.
\end{equation}
There is of course the caveat that the nuclear irradiation probably
occurred at an earlier and very active stage of pre-main-sequence
evolution when the X-ray emission might well have have been
different. While small samples and high obscuration impede thorough
study, CHANDRA observations indicate that the X-ray luminosities and
flares of Class\,I protostars (ages $\sim 0.1$ Ma) are comparable to,
and perhaps even stronger than, the emission from Class \,II and III T
Tauri stars (ages $\sim 0.5-10$ Ma) (Imanishi et al. 2001).  A weak
decline of X-ray emission with age, roughly $L_x \propto t^{-1/3}$, is
seen over the 0.1-10 Ma age range spanned by the COUP stars (Preibisch
\& Feigelson 2005). In any case, all such estimates of particle
fluences are uncertain by factors of several at least.

	In addition to the inferred nuclear spallation effects, the
observed X-rays themselves can directly affect the physical state of
the irradiated material, i.e., the proto-CAIs (calcium-aluminum
inclusions) and chondrules seen in the earliest solar system solids in
chondritic meteorites.  According to Lee et al.~(1998) and Gounelle et
al.~(2002), these solids experienced thermal processing episodes for
several years before being launched into the primitive solar
nebula. These events induced a variety of phase changes, including
partial or full evaporation. Powerful flares, such as those at the
high end of the flare distribution seen by COUP, may play an important
role in the thermal processing of this material because of the larger
energies involved and also because of the enhanced penetrating power
of the hard photons characteristic of flares.

Most of the YSOs observed in the ONC, including the solar-analog
sample of Wolk \etal (2005), are revealed T Tauri stars with a median
age of 2\,Myr. Although a fair fraction have disks and are still
accreting, most of the mass has already been accreted following
earlier more active stages that occurred during the first several
hundred thousand years of their lives. It is very likely that the
nuclear irradiation that led to some if not many of the short-lived
radionuclides occurred in this early period and not during the T Tauri
phase\footnote{The X-ray properties of stars at an earlier stage of
evolution lying deeper inside the Orion molecular cloud have been
studied by Grosso \etal (2005).}.

\section{Summary}

X-ray observations of YSOs with the current generation of large X-ray
telescopes, i.e., CHANDRA and XMM/NEWTON, provide much of the
information needed to study the effects of the X-rays on the formation
of these stars. An extended 10-day observation, the CHANDRA Orion
Ultradeep Project (Getman \etal 2005), has helped clarify our
understanding of the temporal variation of the emission, including the
properties and nature of the flares. Particularly useful for
meteoritics is the study by Wolk \etal (2005) of $\sim 30$ COUP stars
with masses close to solar. This work tells us that the young sun was
luminous in X-rays, emitted many hard photons with energies greater
than 1\,keV, and was susceptible to frequent strong flares. The young
Sun was extraordinarily luminous in X-rays, emitting $\sim 10^{30}$
erg s$^{-1}$ continuously and flares with peak luminosities sometimes
reaching $\sim 10^{31}-10^{32}$ erg s$^{-1}$. The contemporary Sun, in
comparison, typically emits $\simeq 10^{26}-10^{27}$ erg s$^{-1}$ in
the CHANDRA band with occasional flares peaking up to $\simeq 3 \times
10^{28}$ erg s$^{-1}$.  Furthermore, the spectrum of the young Sun was
harder, emitting many photons with energies in the $1-10$ keV range

The X-rays emitted from YSOs can affect all of the near dynamical
components that are involved in the formation of low-mass stars: the
accretion disk, the outflowing wind or winds, the accretion funnel
that actually builds the star, and the magnetically active regions
interior to the inner edge of the accretion disk where magnetic
reconnection, particle acceleration, and spallation reactions occur.
The X-rays help determine the physical and chemical properties of
these regions, and also offer interesting new diagnostic opportunities
for measuring their properties. Particularly interesting is the
possibility that, by determining the level of midplane ionization
important for the MRI, the X-rays affect the accretion that builds the
young star. Equally intriguing is the potential role of the dead zone,
associated with layered accretion, in the formation of planets.
Perhaps most important, the characteristic properties of the X-rays
from sun-like stars provide a foundation for the local irradiation
model for the production of the short-lived radionuclides seen in
chondritic meteorites. These considerations on X-ray interactions
offer considerable promise for the further study of YSO X-rays and
their effects.

%\section{}   %%% Top level section head (remove "%" symbol)
%\subsection{}   %%% Second level section head (remove "%" symbol)
%\subsubsection{}   %%% Lowest level section head (remove "%" symbol)
%\section*{}	%%% Unnumbered top level section head (remove "%" symbol)
%\subsection*{}   %%% Unnumbered second level section head (remove "%" symbol)

%\acknowledgement %%% Text of acknowledgements runs on after this command.

%%% THE BIBLIOGRAPHY
%%%
%%% CONSULT SECTION 3 OF "INSTRUCTIONS FOR AUTHORS" FOR HOW TO USE NATBIB.
%%% AUTHORS ARE ENCOURAGED TO USE EITHER THE "THEBIBLIOGRAPY" ENVIRONMENT
%%% BY UNCOMMENTING (DELETING THE "%" SYMBOL) THE COMMANDS BELOW, OR BY
%%% USING THE BIBTEX ENVIRONMENT. TO FIND OUT WHICH IS APPLICABLE TO YOUR
%%% CONTRIBUTION, CONSULT THE VOLUME EDITORS FOR YOUR PROCEEDINGS.
%%%

\end{document}